# A Maintainability Checklist for Spreadsheets

Henk Vlootman & Felienne Hermans
Vlootman Consultancy and Delft University of Technology
info@vlootman.nl - f.f.j.hermans@tudelft.nl

**ABSTRACT**

*Spreadsheets are widely used in industry, because they are flexible and easy to use. Often, they are even used for business-critical applications. It is however difficult for spreadsheet users to correctly assess the maintainability of spreadsheets. Maintainability of spreadsheets is important, since spreadsheets often have a long lifespan, during which they are used by several users. In this paper, we present a checklist aimed at measuring the maintainability of a spreadsheet. This is achieved via asking several questions, and can indicate whether the spreadsheet is safe to use now and in the future. We demonstrate the applicability of our approach on 11 spreadsheets from the EUSES corpus.*

**1. INTRODUCTION**

Spreadsheets are used extensively in business, for all sorts of tasks and purposes. While other assets of companies---like software products and processes---are strongly guarded, spreadsheets are usually not structurally checked. This lack of control contrasts their impact, which can be widespread, as previous studies have shown. For instance, Hall [Hall, 1996] interviewed 106 spreadsheet developers and found that only 7% of the spreadsheets were of low importance and that as much as 39% were of high importance. In a more recent study we found similar results [Hermans, 2011].

However, these spreadsheets are often of dubious quality, as there are no widely accepted spreadsheet design policies and they are hardly ever enforced or checked. While in the past, research has been conducted on spreadsheet metrics [Bregar, 2004] [Hodnigg, 2008], it remains hard to assess the maintainability of a spreadsheet. In this paper we present a checklist that can be used in industry by spreadsheet professionals to measure a single spreadsheet and to compare different spreadsheets with each other in a fair way. This checklist has been developed by the first author of this paper and has been used to assess over 150 models over the past years.

The checklist aims to measure the maintainability of a given spreadsheet, according to ISO/IEC 9126, this includes ease of understanding, adapting and testing the spreadsheet. Note that we only focus on the maintainability of formulas, other constructs like charts, VBA and pivot tables are out of scope.

In this paper, we describe the rationale behind the checklist and put it to the test, by applying it to 11 randomly selected spreadsheets from a well-known test set and having this analysis done by two different assessors; both authors of this paper.

With this analysis, we want to investigate:

- Whether this checklist is capable of categorizing spreadsheet maintainability
- To what extent the two different assessors obtain similar results
- How well spreadsheets score on the checklist

## 2. THE RATIONALE BEHIND THE CHECKLIST

The checklist consists of several categories, that each contributes to the checklist in a different way. The categories we use are based on existing work in spreadsheets [Read, 1999], [Prior, 2006] combined with on our personal experiences working with spreadsheets in the field.

It is important to note that the current version of the checklist is neither perfect nor complete. In our opinion a checklist is an appropriate way to measure spreadsheet maintainability and the current version is merely a first attempt that has proven valuable in practice.

### 2.1 Documentation

For maintenance and evolution purposes, it is very important that documentation exists [Prior, 2006] Documentation matters for several reasons. The most important is the continuity of the model. If the creator of the model is not present anymore, for whatever reason, problems will occur. And as we saw in previous work, models tend to stay in use for several years. Often it happen that there is just one employee can work with a spreadsheet, and problems occur when this employee is on a holiday.

We distinguish two different types of documentation: technical documentation and user documentation.

*Technical documentation* helps to understand the structure of a model. For instance, it explains why certain design elements were chosen and how chains of formulas are connected. The lack of technical documentation is a serious problem.

*User documentation* helps the user to work with the model. Although not as important as technical documentation, missing or faulty in user documentation can cause the model's users to misuse the model.

### 2.2 Structure

Previous research [Hermans, 2011], [Hermans, 2012a] shows that a clear structure for an Excel model is essential for the understandability of the model. Therefore, spreadsheet structure it is a central category of the checklist. We distinguish three elements of structure: the categorization of worksheets, the naming of worksheets and the separation of calculations and input.

**2.3 Management**

Within control, there are two elements for the checklist. The first one is *changeability*, defined by ISO/IEC9126 as the ease with which we can modify a spreadsheet. Therefore we look at whether a model is easy to understand and whether modifications might influence the results.

The second aspect of the management category is ease of use. When variables and formulas are easy to find, it will be easier to maintain the model.

These two aspects result in questions on the location and visibility of the input cells and the use of named ranges.

**2.4 Safety**

Safety is defined as the robustness of the model: can small changes have large consequences or have control measures against this been implemented? For this category we look at the normalization of the variables. With this we mean that if how constants in the model (i.e. Value Added Tax) how are stored on the spreadsheet and how to formulas utilizing this constant reference to it. Furthermore, we look at the use of user selection choices. Are control elements used and are inputs validated?

**2.5 Formatting**

Good formatting can support the user, this is the reason that we added this as a category is our check list. In this category, we look at how cells are formatted and whether the creator of the model added comment boxes or explanation.

**2.6 Skills**

In the final category, we aim to measure the skills of the creator of the spreadsheet. The more complex Excel elements are used, the higher the skills of the user is ranked. This method is based on the previous work of [Hole, 2009]

**3. THE CHECKLIST**

This section contains the questions of the checklist. This checklist can also be found online: http://www.vlootman.nl/index.php?option=com_weblinks&view=category&id=12:bestanden&Itemid=30&lang=en

**3.1 Weighting the questions**

In Section 3.2, the weightings for each category are given in brackets. This weighing has been chosen based on personal experience of working with the checklist in industry. However, we do not consider the specific weights given to a question as a part of the checklist model. Rather, we open up the possibility for users to calibrate these weights to represent their own assertion of the value of each question. As long as the weights are kept constant while measuring multiple spreadsheets, the checklist is suited to compare different spreadsheets.

**3.2 The questions**

**Documentation**

1) Is there any technical description available? 20
2) Is there any user description available? 15

**Structure**

3) Are the sheets grouped by function? 20
4) Is the naming of the worksheets understandable? 5
5) Are calculations separated from input? 15

**Management**

6) Are all variables placed together? 10
7) Is there a clear distinction between input and output? 10
8) The output is compact and clear? 10
9) Are valid Excel ranges used? 10
10) Are the input cells logically grouped? 10

**Safety**

11) Is normalization used on the variables? 20
12) In which way will the user selection be processed? 15

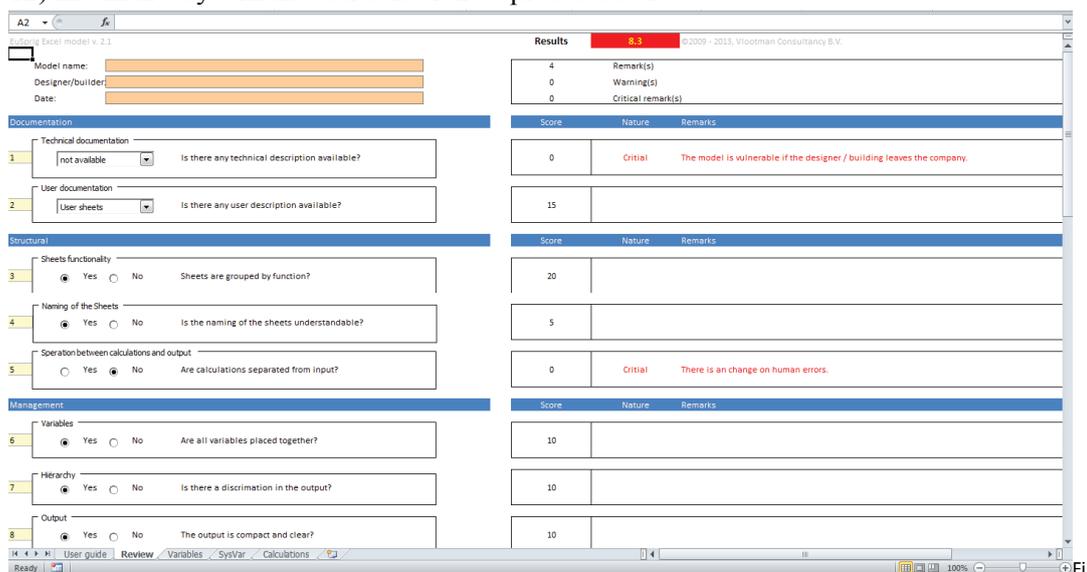

Figure 1 A screenshot of the check list in Excel

**Formatting**

13) Are the input cells formatted consequently? 10
14) Are the output cells formatted consequently? 10
15) Are the other cells formatted consequently? 10
16) In which way will the user be supported? 15

**Skills**

17) Is array functionality used in the model? 20
18) Does the model support windows for the output? 15
19) Are names used in the model? 15
20) Are names separated in categories? 10
21) Are the names consistently composed? 5
22) Are the names consistently used? 5
23) Are (complex) single sided functions used in the model? 10
24) Does the model have nested functions? 15
25) Does the model have links towards other cells? 10
26) Does the model have absolute links or names towards other cells? 5

## 4 EVALUATION

### 4.1 Goal and setup

To evaluate our checklist; we performed an empirical study in which both authors checked 11 spreadsheets from a well-known test set of spreadsheets.

With these experiments we want to validate that

- The checklist distinguishes spreadsheets from each other
- The results can be used to assess a spreadsheet
- The results can be used to improve a spreadsheet

### 4.2 Data

As a source for the experimental phase we used spreadsheets from the EUSES Spreadsheet Corpus [Fisher, 2005]. This corpus consists of around 5000 spreadsheets, divided over 11 categories---ranging from educational to financial---and has been used by several researchers to evaluate algorithms on spreadsheets, among which [Hermans, 2010] and [Abraham, 2006]. We randomly selected one spreadsheet from each of the 11 categories. All 11 spreadsheets can be downloaded from our research page
(http://swerl.tudelft.nl/bin/view/FelienneHermans/).

## 4.3 Results

Table 1 Results of the check list on the EUSES corpus by the first author. Highlighted cells are questions where authors disagreed

| Q | 1 | 2 | 3 | 4 | 5 | 6 | 7 | 8 | 9 | 10 | 11 |
|---|---|---|---|---|---|---|---|---|---|---|---|
| **Documentation** | | | | | | | | | | | |
| 1 | N/A | N/A | N/A | N/A | N/A | N/A | N/A | N/A | N/A | N/A | N/A |
| 2 | N/A | N/A | N/A | N/A | N/A | N/A | N/A | N/A | User sheets | N/A | N/A |
| **Structure** | | | | | | | | | | | |
| 3 | No | No | Yes | No | No | No | No | No | No | No | No |
| 4 | Yes | Yes | Yes | Yes | Yes | No | No | Yes | No | No | Yes |
| 5 | No | No | Yes | No | No | No | No | No | No | No | No |
| **Management** | | | | | | | | | | | |
| 6 | No | Yes | Yes | Yes | No | No | No | No | Yes | No | No |
| 7 | Yes | Yes | Yes | Yes | No | Yes | Yes | No | No | No | No |
| 8 | Yes | Yes | Yes | Yes | Yes | Yes | Yes | Yes | Yes | Yes | Yes |
| 9 | Yes | Yes | Yes | Yes | Yes | Yes | Yes | Yes | Yes | Yes | Yes |
| 10 | Yes | No | Yes | N/A | Yes | N/A | No | No | Yes | Yes | Yes |
| **Safety** | | | | | | | | | | | |
| 11 | No | No | No | No | No | No | No | No | No | No | No |
| 12 | No | No | No | Controls | No | No | No | No | No | No | No |
| **Formatting** | | | | | | | | | | | |
| 13 | No | No | No | No | No | No | Yes | No | No | No | No |
| 14 | No | No | No | No | No | No | Yes | No | No | No | No |
| 15 | No | No | No | No | No | No | No | No | No | No | No |
| 16 | Not | Not | Not | Not | Not | Not | In cells | Not | Not | Not | Not |
| **Skills** | | | | | | | | | | | |
| 17 | No | No | No | Yes | No | No | No | No | No | No | No |
| 18 | No | No | No | No | No | No | No | No | No | No | No |
| 19 | No | No | No | No | No | No | No | No | No | No | No |
| 20 | No | No | No | No | No | No | No | No | No | No | No |
| 21 | No | No | No | No | No | No | No | No | No | No | No |
| 22 | No | No | No | No | No | No | No | No | No | No | No |
| 23 | No | No | Yes | Yes | No | Yes | No | Yes | No | No | Yes |
| 24 | No | No | Yes | Yes | No | Yes | No | Yes | No | No | No |
| 25 | Yes | Yes | Yes | Yes | Yes | Yes | Yes | Yes | Yes | Yes | Yes |
| 26 | No | No | Yes | Yes | No | Yes | No | Yes | Yes | No | Yes |

**Table 2 Scores corresponding to the answers found in Table 1, as calculated by the current weighting in the checklist**

| Model | Doc. | Structure | Man. | Safety | Format | Skills | Overall |
|---|---|---|---|---|---|---|---|
| Posey_Q | 0 | 0 | 6 | 0 | 0 | 0.9 | 1.3 |
| FinalBudget | 0 | 1.3 | 6 | 0 | 0 | 0.9 | 1.4 |
| CHOFAS | 0 | 10 | 10 | 0 | 0 | 3.6 | 4.1 |
| FinFun | 0 | 0 | 8 | 0 | 0 | 3.6 | 2.5 |
| karen-cs101gradebookSp98 | 0 | 1.3 | 4 | 0 | 0 | 3.6 | 2.1 |
| 9-Grade | 0 | 1.3 | 8 | 0 | 0 | 0.9 | 1.7 |
| Solutions_week_3 | 0 | 1.3 | 6 | 0 | 0 | 2.3 | 1.9 |
| grain_inventory_market#A85C5 | 0 | 0 | 6 | 0 | 6.7 | 0.9 | 2.2 |
| Equity2 | 0 | 1.3 | 10 | 4.3 | 0 | 5.5 | 4.1 |
| occupancy_schedules_m#A82E1 | 4.3 | 0 | 8 | 0 | 0 | 1.4 | 2.2 |
| BTVSCCG Inventory | 0 | 1.3 | 8 | 0 | 0 | 0.9 | 1.7 |

**Difference between assessors**

The first observation that can be made in Table 1, is the fact that the results of both authors are not entirely consistent. Even in the more technical categories, like skills, we see differences, like on the issue of nested functions (Q24) Here the discrepancy is whether a formula F14*(1-F16) is or is not a nested function. Author 1 only considers built-in functions (like SUM) to count as nested, where author 2 also counted infix operations like as functions. These results indicate that more guidance in answering the questions would improve consistent scoring.

**Documentation**

The second fact that stands out is the fact that all but 1 spreadsheets miss both technical and users documentation. This might be because for these spreadsheets documentation exists outside of the spreadsheets. This would not have been gathered by the creators of the EUSES, which only searches for spreadsheets. However, we do not believe this is the case for all of those spreadsheets, we think that these results show that the general state of documentation is poor in spreadsheets.

**Structure**

Within the structure category, it is interesting to see that in one of the cases, the assessor differ in their opinion on clear naming of worksheets. In this case (file 8) there are some worksheets with standard names (sheet 13, sheet 14, etc.) and some other where a name has been changed by the user. In the majority of the cases (7 spreadsheets) clear names are used. However, worksheets are never clearly designated for input of output, in all models this is mixed over worksheets.

**Management**

Management is the category that overall scores the highest of all, with scores from 4 to the maximum value of 10. This is mainly due to the fact that the spreadsheets clearly separate input and output quite well, that output is grouped and that valid ranges are used. Probably, the management category scores high, as spreadsheet users have to be able to find their output easy, so arranging this wisely is something they do, as opposed to building in safety constraints, which is not needed for their daily tasks.

**Safety**

In the safety category too, the scores are low. This means that often, hardcoded values are used. In only one of the spreadsheets, user controls were introduced to prevent erroneous and validations were not used in one single test spreadsheet.

**Formatting**

In the formatting category too, there is only one spreadsheet scoring over 0, this is file 7, where input and output cells are consistently colored and the user gets help with their input too. In this category too, there are some differences between the assessors, mainly on the question whether non-input and output cells, i.e. labels, are consistently formatted.

**Skills**

The skills category is a diverse category, with scores ranging from 0.9 to 5.5, as can be seen in Table 2. As mentioned before, there are differences in judgement of skill between authors as the nature of this work is highly technical. Further, the results might be biased since 4 questions in this category concern naming,. Amodel lacking any names will score very low on skills because of this. In retrospect, this focus might be a bit heavy.

4.4 Conclusion

The aim of this study was to validate that a spreadsheet checklist can help users to compare spreadsheets in an objective way.

Based on the results we have presented above, we can conclude that randomly chosen spreadsheets from the EUSES corpus differ on many aspects of the checklist, confirming the idea that this checklist helps to distinguish between spreadsheets. Furthermore we observe that although there are some differences between the two assessors, overall the results are consistent enough to conclude that the checklist can be used to compare spreadsheets in a fair way, even if assessed by different people. Finally, we see that the spreadsheets under observation in this study do not score high, which is consistent with experiences from the first author while using this checklist in practice: many spreadsheets could be greatly improved

## 5. DISCUSSION

The current version of the checklist is able to measure the maintainability of a spreadsheet, in other words, to measure how easy to understand and change the spreadsheet will be for future users. However, there are some limitations. For instance, in the current version of the checklist only focuses on formulas and hence graphs, charts and pivot tables are not taken into account. The following subsections describe other avenues for improvement.

### 5.1 Automating the checklist

Some of the categories of the check list, such as the skills category can be easily automated. This would reduce the work load for the assessor. We do however prefer a partly automated system over a fully automated, as some categories like documentation can never be automated and also, using a fully automated system might make it easy for spreadsheet users to game the metrics.

### 5.2 Taking smells into account

In previous research [Hermans, 2012a; Hermans, 2012b] we have worked on the detection of *smells* in spreadsheets. In that work we have found the 'smelly' formulas can diminish maintainability of a spreadsheet. Hence, it would be interesting to add a category specific for 'smells'. Also, spreadsheet complexity metrics, like the ones suggested in [Bregar, 2004] or [Hodnigg, 2008] could make our assessment method more precise.

### 5.3 Suggesting improvements

Many of the questions of the check list are actually guidelines in disguise, such as the use of naming or a clear difference in lay-out between input and output cells. Having a low mark on one of these questions, might be caused by the fact that a spreadsheet user in unaware of these underlying guidelines. Explaining these guidelines and even giving concrete help on how to improve-such as 'move these and these formula cells to worksheet 'Sheet1'- could help users improve their score.

### 5.4 Making ranges continuous

In the current version of the checklist, many questions are dichotomous yes/no and in when using the checklist, we found that sometimes an option in between would make assessing more fair. Another possibility would be to add an option for 'not applicable'.

## 6 CONCLUDING REMARKS

The aim of this paper is to develop a set of questions that indicate the maintainability of a given spreadsheet. To that end we have suggested such a checklist and applied in on 11 spreadsheets from the EUSES corpus. We conclude that, while this paper represents a first attempt, our checklist is capable of differentiating between different spreadsheets and identifying areas for improvement.

The key contributions of this paper are as follows:

- A comprehensive checklist to assess spreadsheet maintainability
- An empirical evaluation investigating the usefulness of that checklist

The current research gives rise to several directions for future work. Firstly, it would be very interesting to perform a more thorough empirical evaluation of this checklist to validate whether it can prevent errors or problems in practice. Furthermore, it would be interesting to couple low scores to concrete actions, which help users to improve their spreadsheets. This could be combined with the above proposed automation of the checklist.